\documentclass[11pt,twoside]{article}
\usepackage{asp2010}
\usepackage{hyperref}
\usepackage{graphicx}
\usepackage{natbib}
\usepackage{wrapfig}

\resetcounters

\begin{document}

\title{Integral-field spectroscopy of the young high-mass star IRAS13481-6124}
\author{Bringfried Stecklum, $^1$ Alessio Caratti o Garatti, $^2$ and Hendrik Linz $^3$ 
\affil{$^1$Th\"uringer Landessternwarte, Tautenburg, Germany}
\affil{$^2$Max-Planck Institut f\"ur Radioastronomie, Bonn, Germany}
\affil{$^3$Max-Planck Institut f\"ur Astronomie, Heidelberg, Germany}
}
 
\begin{abstract}
We present results of AO-assisted K-band IFU spectroscopy of the massive young star IRAS13481-6124 performed with ESO's VLT/SINFONI instrument. Our spectro-astrometric analysis of the Br$\gamma$ line revealed a photo-center shift with respect to the adjacent continuum of $\sim$1\,AU at a distance of 3.1\,kpc. The position angle of this shift matches with that of the outflow which confirms that the massive star is indeed the driving source. Furthermore, a velocity gradient along the major disk axis was found which hints at the rotational sense of the ionized region, and thus of the disk as well. The gradient is not consistent with Keplerian motion but points to rigid rotation of the innermost disk.
Notably, emission of H$_2$ is absent from source while both shocked and fluorescent H$_2$ emission are observed in its immediate surroundings.

\end{abstract}

\section{Introduction}
 IRAS13481-6124 was considered as potential massive young stellar object (MYSO) by Chan ea. (1996\nocite{1996A&AS..115..285C}). SED modeling confirmed this view by suggesting a stellar mass of $\sim$20\,$M_\odot$ (Grave \& Kumar 2009\nocite{2009A&A...498..147G}). Recently, radio-interferometric molecular line observations as well as H$_2$ narrow-band imaging showed that it drives a parsec-scale bi-polar outflow, and VLTI measurements in the near- and thermal-infrared revealed the presence of a circumstellar disk (Kraus ea. 2010\nocite{2010Natur.466..339K}, Stecklum ea. 2010\nocite{2010hmsf.conf}). At the distance of 3.1$\pm$1.1\,kpc (Urquhart ea. 2007\nocite{2007A&A...474..891U}) milli-arcsecond (mas) resolution is required to obtain information on the spatio-kinematics of the inner disk from emission lines, e.g. Br$\gamma$. Attempts to retrieve this with the VLTI failed so far because of sensitivity issues. Therefore, we have applied an alternative approach based on spectro-astrometry (SA).

SA aims at measuring the photo-center shift as a function of wavelength, and goes back to the concept of differential speckle interferometry (Beckers 1982\nocite{1982AcOpt..29..361B}). Since the accuracy is $\propto$SNR, SA demands bright targets. The first successful attempt (Bailey 1998\nocite{1998MNRAS.301..161B}) utilized long-slit spectroscopy at two position angles offset by 90\deg{} to reveal SA shifts in emission lines using the adjacent continuum as reference. Thereby, evidence for companions and outflows was found. Although the complemental use of two slit orientations (original and reversed) helps canceling instrumental effects, systematics still exist (Brannigan ea. 2006\nocite{2006MNRAS.367..315B}). The SA analysis of IFU spectroscopy profits from the true 2D observing nature, and benefits from Strehl improvements when coupled with AO (Davies ea. 2010\nocite{2010MNRAS.402.1504D}). This boosts the SA accuracy into the $\mu$as range (Goto ea. 2012\nocite{2012ApJ...748....6G}).

\section{Observations and data reduction}
K-band (R$\approx$2000) IFU spectroscopy of IRAS13481-6124 was obtained with ESO's SINFONI instrument in April 2011 using a natural guide star 16\arcsec{} NW of the target as wavefront reference. A dither map covering the inner flow region but excluding the bright (Ks$=$4.9) MYSO to avoid detector remanence was taken at the 0\farcs25 pixel scale (cf. Fig.\ref{fig1}).  The MYSO itself was observed with the 0\farcs025 pixel scale at the total on-source time of 600\,s. The results presented below are based on the pipeline-reduced data. The SA analysis follows from Goto ea. (2012)\nocite{2012ApJ...748....6G}. Centroid positions [x,y] for each wavelength slice of the data cube were obtained by fitting a generalized 2D-Gaussian to the PSF for the range from 2.147\,$\mu$m to 2.185\,$\mu$m. The instrumentally induced positional bias was removed by a polynomial fit to the x and y dependence on wavelength which excluded the Br$\gamma$ line. The 2nd order moment, i.e. the PSF width, was normalized to the continuum to account for the wavelength dependence. 

\section{Results and discussion}

\begin{wrapfigure}{r}{0.5\textwidth}
  \vspace*{-.35cm}
   \includegraphics[clip=true,width=0.49\textwidth]{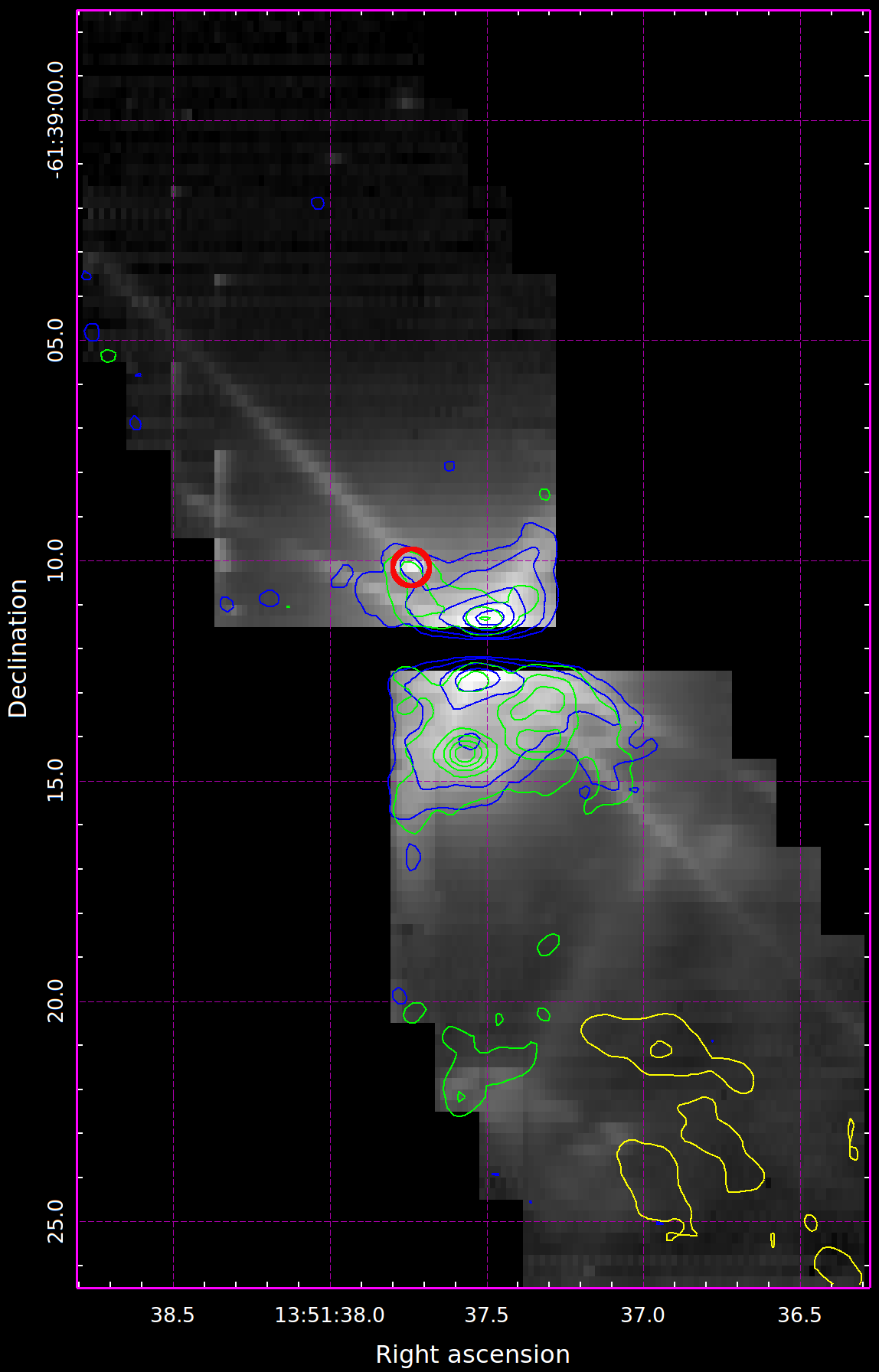}
   \caption{K-band mosaic at 0\farcs25 pixel scale. }
\label{fig1}
\end{wrapfigure}

The immediate environment of the MYSO is shown in Fig.\ref{fig1} together with contours of Br$\gamma$ (blue) as well as H$_2$ emission. The long narrow spikes are diffraction
artifacts. The mosaic does not reveal any emission from the red-shifted lobe to the north-east but discloses the faint companion at 2\farcs5 distance (red circle) already detected in the N-band (Stecklum ea. 2010\nocite{2010hmsf.conf}).  The blue-shifted lobe to the south-west is associated with scattered continuum as well as H$_2$ emission. The molecular hydrogen is excited by fluorescence (green) and shocks (yellow) as inferred by the 2$-$1\,S(1)/1$-$0\,S(1) line ratio. A blob situated 2\arcsec{} south of the MYSO is particularly bright in Br$\gamma$ as well fluorescent H$_2$. It is likely a PDR, caused by the intense UV illumination of the nearby MYSO. Notably, the highly collimated H$_2$ flow cannot be traced down to the driving source. This is confirmed by the lack of H$_2$ lines on the central target in the very high resolution data.

\begin{figure}[!h]
  \vspace*{-.35cm}
   \includegraphics[clip=true,width=6cm]{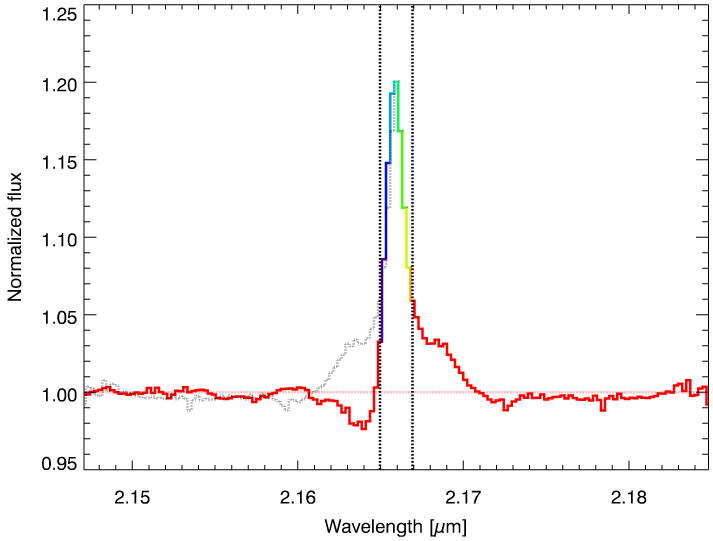}
   \includegraphics[clip=true,width=6cm]{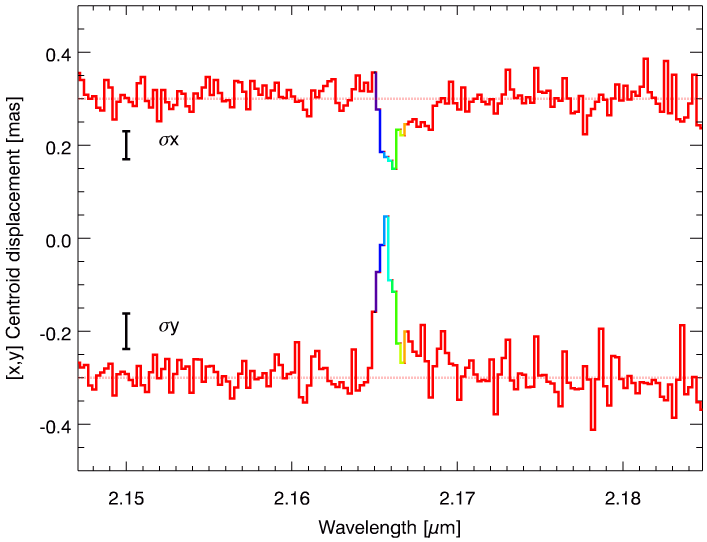}
   \caption{{\it Left: } Br$\gamma$ line and adjacent continuum
                 {\it Right:} SA signal vs. wavelength 
                }
   \label{fig2}
\end{figure}

The Br$\gamma$ line seen in the latter (Fig.\,\ref{fig2}\, left) shows a shallow P Cygni profile, and evidence for high-velocity wings from an ionized wind.  The optically thick wind was found already in the recombination line study of Beck ea. (1991\nocite{1991ApJ...383..336B}) but their SNR precluded the detection of the Br$\gamma$ P Cygni profile. In order to illustrate the absorption trough as well as the wings, the spectrum longward of the peak was flipped, and plotted in gray. For the SA analysis only the core of the line (enclosed by the vertical lines), unaffected by absorption, was used. Fig.\,\ref{fig2} (right) displays the results where the vertical bars denote the 1\,$\sigma$ error which has a mean of 45\,$\mu$as. SA signals are clearly detected across the line in the [x,y] centroids. The SA accuracy is in good agreement with that of Goto ea. (2012)\nocite{2012ApJ...748....6G} which confirms the well-behaved systematics.

\begin{wrapfigure}{r}{0.58\textwidth}
  \vspace*{-.25cm}
   \includegraphics[clip=true,width=0.57\textwidth]{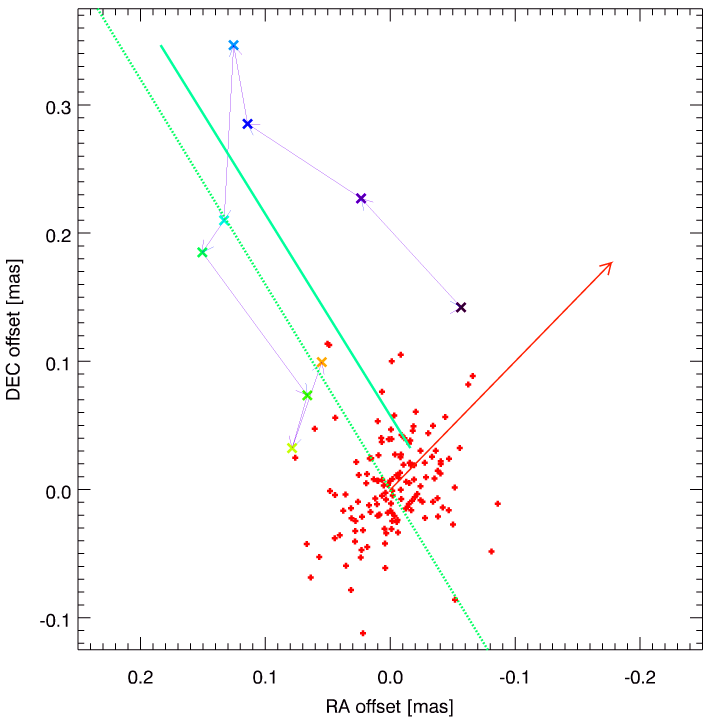}
   \caption{SA map showing the Br$\gamma$ line excursion.}
   \label{fig3}
\end{wrapfigure}
\vspace*{-.15cm}

In order to put these results into relation with the disk and the outflow, they were plotted in the SA map (Fig.\,\ref{fig3}). Red crosses mark continuum while the line core is color coded, and connected by thin arrows. The ellipticity of the continuum distribution is due to anisoplanatism (the red arrow points along the direction to the wavefront reference star). The long light-green line indicates the PA (32\deg) of the parsec-scale outflow. The thick green line is the major axis of the SA excursion at the PA of 29\deg. The interpretation of this map requires awareness of the location of the continuum. To this aim, high-resolution synthetic images were generated for best-fitting SED models (Robitaille ea. 2007\nocite{2007ApJS..169..328R}) using the radiative transfer code of Whitney ea. (2003\nocite{2003ApJ...591.1049W}), and analyzed in the same way as the real data. This indicates that the continuum centroid is in between the star and the south-western disk rim. Therefore, the Br$\gamma$ excursion is pointing towards the star, i.e. the ionized gas is more confined. The fact that the position angle of the Br$\gamma$ excursion matches with that of the large-scale outflow confirms that the MYSO  is the driving source of the outflow indeed. 

The SA analysis also yielded marginal signals in the PSF width, indicating that the line-emitting region is more compact than the continuum. This corresponds to previous results for Herbig Ae/Be stars obtained with VLTI/AMBER (Kraus ea. 2008\nocite{2008A&A...489.1157K}).  On the contrary, the self-absorption region appears to be slightly more extended. 

\begin{wrapfigure}{l}{0.5\textwidth}
   \includegraphics[clip=true,width=0.48\textwidth]{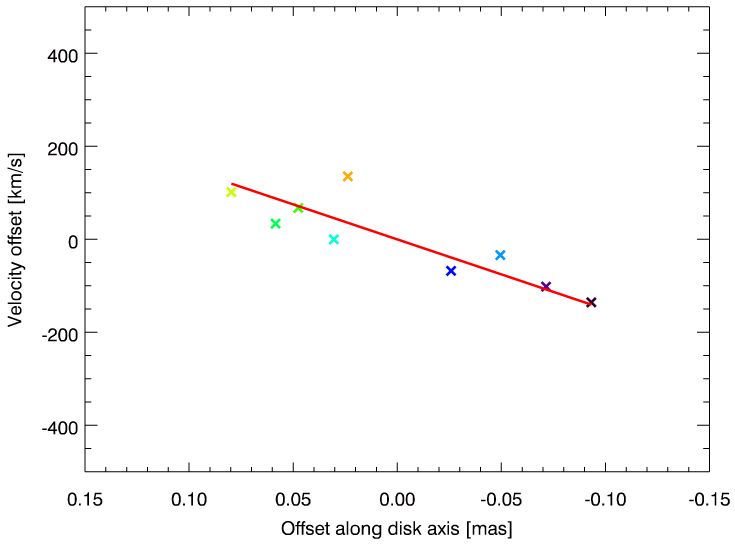}
   \caption{ Velocity gradient along the major disk axis}
   \label{fig4}
\end{wrapfigure}

From the SA map it can be noticed that the velocities left- and right-ward of the major axis of the SA excursion differ systematically. The spatio-kinematics corresponds to a velocity gradient along the major disk axis (Fig.\,\ref{fig4}). It is inconsistent with Keplerian motion, and hints at a rigid rotation of the innermost disk, possibly due to a magnetic confinement.  According to the gradient, the eastern part of the disk is receding.
 
The attempt to apply the SA analysis to the continuum as well was spoiled by the lower accuracy of the SA signals of the fainter standard star. However, given the high SA accuracy which can be achieved by IFU spectroscopy of bright YSOs, tracing the wavelength-induced wander of the continuum centroid seems possible using equally bright, nearby comparison stars. Thereby, constraints on inclination and source extent could be achieved in the absence or as a complement of time-demanding interferometric observations.

\acknowledgements
Based on observations made with ESO Telescopes at the La Silla Paranal Observatory under program ID 087.C-0951(B). Thanks to ESO staff,  especially D. N\"urnberger, for obtaining the requested data eventually. 

\bibliography{p23_stecklum}

\end{document}